\documentclass[twocolumn]{revtex4-1}
\usepackage{amssymb,epsfig,amscd}
\usepackage{graphicx}% Include figure files
\usepackage{dcolumn}% Align table columns on decimal point
\usepackage{bm}% bold math

\newcommand{\dirac}{\partial\llap{$\diagup$\kern-2pt}}

\def\be{\begin{equation}} 
\def\ee{\end{equation}}
\def\bq{\begin{eqnarray}} 
\def\eq{\end{eqnarray}}

\begin{document}

\title{Magnetic field generated by r-modes in accreting quark stars}
\author{L.~Bonanno$^b$, C.~Cuofano$^a$, A.~Drago$^a$, G.~Pagliara$^b$, J.~Schaffner-Bielich$^b$}

\affiliation{$^a$ Dipartimento di Fisica, Universit\'a di Ferrara and INFN sezione di Ferrara, 44100 Ferrara, Italy\\ 
$^b$Institut f\"{u}r Theoretische Physik, Ruprecht-Karls-Universit\"at,
Philosophenweg 16,  D-69120, Heidelberg, Germany}

\begin{abstract}
We show that the r-mode instability can generate strong toroidal fields in the core of 
accreting millisecond quark stars by inducing differential rotation. We follow the 
spin frequency evolution on a long time scale taking into account the magnetic damping rate in the 
evolution equations of r-modes. The maximum spin frequency of the star is only marginally smaller
than in the absence of the magnetic field. The late-time evolution of the stars which enter the
r-mode instability region is instead rather different if the generated magnetic fields are taken
into account: they leave the millisecond pulsar region and they become radio pulsars. 
\end{abstract}

\maketitle

\section{Introduction}
Since the first paper in which r-modes in rotating neutron stars were
shown to be unstable with respect to the emission of gravitational
waves \cite{Andersson:1997xt}, it was recognized their relevance
in Astrophysics to explain the observed distribution of the rotation
frequency of stars in Low-Mass-X-Ray-Binaries (LMXBs) (see
\cite{Andersson:2000mf} for a review). On the other hand, the
instability triggered by r-modes is actually damped to some extent by
the viscosity of the matter composing the star: the higher the
viscosity, the higher is the frequency of rotation of the star. This
fact opens the possibility of investigating the internal composition of
compact stars by studying the viscosity of the different possible high
density phases which can appear in these stellar objects. A number of
papers are currently present in the literature about the shear and
bulk viscosities of nucleonic matter
\cite{Sawyer:1989dp,Haensel:1992zz,Haensel:2000vz,Haensel:2001mw,Benhar:2007yj},
hyperonic
matter\cite{Lindblom:2001hd,Haensel:2001em,Chatterjee:2006hy,Chatterjee:2007iw,Gusakov:2008hv,Sinha:2008wb,Jha:2010an},
kaon condensed matter \cite{Chatterjee:2007qs,Chatterjee:2007ka}, pure
quark phases
\cite{Madsen:1992sx,Madsen:1999ci,Alford:2006gy,Sa'd:2006qv,Sa'd:2007ud,Alford:2007rw,Blaschke:2007bv,Sa'd:2008gf,Mannarelli:2008je,Jaikumar:2008kh,Alford:2009jm},
and mixed phases \cite{Drago:2003wg}. 
Interestingly, by studying the so called ``window of instability'' of the r-modes, which is determined 
by the shear and the bulk viscosity of the matter, it was pointed out in \cite{Drago:2007iy} that 
a detection of a sub-millisecond rotating star would indicate the existence of a very viscous phase in the star
with only quark or hybrid stars as possible candidates. 

Also the temporal evolution of the spin frequency of a star under the
effect of r-modes instability has been studied for different types of
composition and different physical systems: newly born compact stars
and old stars in binaries
\cite{Lindblom:1998wf,Owen:1998xg,Andersson:2001ev,Wagoner:2002vr,Drago:2004nx,Drago:2007iy}.
While neutron stars could be very powerful gravitational waves
emitter only during their first months after birth
\cite{Owen:1998xg}, hyperonic and quark or hybrid stars could turn
into steady gravitational waves sources if present in LMXBs
\cite{Andersson:2001ev,Wagoner:2002vr,Reisenegger:2003cq}.

R-modes are also responsible for differential rotation in the star
which in turn generates a toroidal magnetic field: besides the
viscosity of matter, the production of this magnetic field represents
a very efficient damping mechanism for the r-modes. This effect has
been proposed and investigated in
\cite{Rezzolla2000ApJ,Rezzolla:2001di,Rezzolla:2001dh} for the case of a newly born
neutron star and it has been recently included within the r-mode equations 
of the neutron stars in LMXB \cite{Cuofano:2009xy,Cuofano:2009yg}: by calculating 
the back-reaction of the magnetic field on the r-modes instability, 
it has been proven that magnetic fields of the order of $10^{15}$ G can be
produced. Remarkably, this mechanism could be at the origin of the
enormous magnetic field of magnetars.
  
In this paper we extend the calculations of
\cite{Cuofano:2009xy,Cuofano:2009yg} to the case of quark stars in
LMXB. The motivation for this investigation is that the evolution of
accreting stars and their internal magnetic field depends strongly on
the r-modes instability window which, for quark stars, is
qualitatively different from the one of neutron star. Indeed, due to
the large contribution to the bulk viscosity of the non leptonic weak
decays occurring in strange quark matter, the window of instability
splits into two windows, a small one at large temperatures (which is
actually irrelevant for the evolution) and a big one at low
temperatures \cite{Madsen:1999ci,Drago:2007iy}.  Moreover, quark stars
do not have a crust (or just a very thin crust) and, as we will
discuss, this prevents the trapping of the internal magnetic field
developed during the evolution with possible observable signatures.

The paper is organized as follows: in Sec.~II we introduce the system of equations
which provide the temporal evolution of a compact star in a LMXB. In Sec.~III
we show the results of our numerical calculations and finally in Sec.~IV we present 
of conclusions.

\section{The temporal evolution model}
\subsection{R-modes equations}
Let us review the derivation of the equations for the evolution of a compact star in presence of the r-modes.
\\
Following the discussion of Ref.~\cite{Wagoner:2002vr}, we start from the conservation of the angular momentum. 
The total angular momentum of the star can be decomposed into an equilibrium angular momentum $J_*$
and a canonical angular momentum $J_c$, proportional to the r-modes amplitude $\alpha$:
\begin{eqnarray}
J_{tot}=J_*+(1-K_j)J_c\nonumber\\
J_c=-K_c \alpha^2 J_*\label{angcons}
\end{eqnarray}
where $J_*=I_*\Omega$, with $\Omega$ the angular velocity and $I_*=\tilde I M R^2$ the moment of inertia of a star of mass $M$
and radius $R$. For a n=1 polytrope $\tilde I=0.261$, $K_c=9.4\times 10^{-2}$ (see Ref.~\cite{Owen:1998xg}) 
and $K_j$ is of the order of unity (the results of our calculations are rather 
insensitive to the value of $K_j$).

The equations for the evolution of a star are based on two simple considerations:
\begin{itemize}
\item the canonical angular momentum, proportional to the r-modes amplitude, increases 
with the emission of gravitational waves and decreases due the presence of damping mechanisms (as e.g. viscosity and the torque due to an internal magnetic field $B$):
\begin{eqnarray}
\frac{dJ_c}{dt}&=&2 J_c \big(\frac{1}{\tau_{GW}(M,\Omega)}\nonumber\\
&-&\frac{1}{\tau_{damp}(M,\Omega,T,B)}\big)\label{jcan}
\end{eqnarray}
where $\tau_{GW}(M,\Omega)$ and $\tau_{damp}(M,\Omega,T,B)$ are the gravitational wave emission timescale
and the damping timescale, respectively, where the latter includes all the mechanisms muffling the r-modes.
Finally with $T$ we mean the average temperature of the star.
\item the total angular momentum takes contributions from the mass accretion and decreases due to the emission of
gravitational waves and electromagnetic waves. The electromagnetic wave emission is associated to the presence 
of an external poloidal magnetic field which is not aligned with the rotation axis. Indicating with $\tau_{m_e}$ the braking timescale
due to the external magnetic field, the second equation reads:
\begin{eqnarray}
\frac{dJ_{tot}}{dt}=\frac{2 J_c}{\tau_{GW}}+ \dot J_a - \frac{J_*}{\tau_{m_e}}\label{jtot}
\end{eqnarray}
where $\dot J_a$ is the variation of the angular momentum due to the mass accretion. Following Ref.~\cite{Andersson:2001ev}, we assume 
$\dot J_a=\dot M (G M R)^{1/2}$.
\end{itemize}

Combining together the equations (\ref{angcons}), (\ref{jcan}) and (\ref{jtot}), it is easy to obtain the evolution equations
for $\Omega$ and for the r-modes amplitude $\alpha$:
\begin{eqnarray}
\frac{d\alpha}{dt}&=&\alpha\big(\frac{1}{\tau_{GW}}-\frac{1}{\tau_{damp}}\big)\nonumber\\
&+&K_c \alpha^3\big[K_j\frac{1}{\tau_{GW}}+(1-K_j)\frac{1}{\tau_{damp}}\big]\nonumber\\
&-&\frac{\alpha \dot M}{2 \tilde I \Omega}\big(\frac{G}{M R^3}\big)^{1/2}+\frac{\alpha}{2 \tau_{m_e}} \label{eq4} \\
\frac{d\Omega}{dt}&=&-2 K_c\Omega\alpha^3\big[K_j\frac{1}{\tau_{GW}}+(1-K_j)\frac{1}{\tau_{damp}}\big]\nonumber\\
&-&\frac{\dot M\Omega}{M}+\frac{\dot M}{\tilde I}\big(\frac{G}{M R^3}\big)^{1/2}-\frac{\Omega}{\tau_{m_e}}
\label{eq5}
\end{eqnarray}

Here we adopt the estimate given in Ref.~\cite{Andersson:2000mf} for the gravitational radiation rate due to the $l=m=2$ current multipole:
\begin{eqnarray}
\frac{1}{\tau_{GW}}=\frac{1}{47}M_{1.4}R_{10}^4 P_{-3}^{-6}\quad s^{-1}
\end{eqnarray}
where we have used the notation $M_{1.4}=M/1.4 M_{\odot}$, $R_{10}=R/10$ km and $P_{-3}=P/1$ ms.

Following Ref.~\cite{Cuofano:2009yg}, the total damping rate is given by the sum of the shear viscosity and bulk viscosity rates
plus the damping rate due to the internal magnetic field:
\begin{eqnarray}
\tau_{damp}^{-1}=\tau_{s}^{-1}+\tau_{b}^{-1}+\tau_{m_i}^{-1}
\end{eqnarray}
The estimates of the viscosity damping rates for the case of pure quark matter are given in Ref.~\cite{Andersson:2001ev}.
For the shear viscosity it is found:
\begin{eqnarray}
\tau_{s}=3.4\times 10^9 \alpha_s^{5/3}M_{1.4}^{-5/9}R_{10}^{11/3} T_9^{5/3}\quad s\label{svis}
\end{eqnarray}
\\
where $\alpha_s$ is the strong coupling constant and $T_9=T/10^9$ K.
For the bulk viscosity, the viscosity coefficient takes the form
\begin{eqnarray}
\zeta=\frac{\tilde\alpha T^2}{\omega^2+\beta T^4}
\end{eqnarray}
where $\omega$ is the frequency of the r-modes in the corotating frame and $\tilde\alpha$ and $\beta$ are coefficients given in Ref.\cite{Madsen:1992sx}.
Due to this behavior, bulk viscosity is very large at $T\sim 10^9 K$ and gets weaker at 
both lower and higher temperatures. The instability window is then splitted into
a Low Temperature Window (LTW) (at $T<10^9$ K) and an High Temperature Window (HTW) ($T>10^9$ K). Due to the fast cooling rate of quark stars, the (HTW) is crossed very 
rapidly and does not have a big impact on the evolution of the star.
So we are interested only in the LTW. The bulk viscosity damping timescale in the low temperature limit ($T<10^9$ K) is given by\cite{Andersson:2001ev}:
\begin{eqnarray}
\tau_{b}^{low}=7.9 M_{1.4}^2 R_{10}^{-4} P_{-3}^2 T_9^{-2} m_{100}^{-4}\quad s \label{bvis}
\end{eqnarray}
where $m_{100}$ is the mass of the strange quark in units of 100 MeV. Notice that the value bulk viscosity damping timescale is strongly dependent on the value of $m_{100}$.\\
\\
The expression of the magnetic damping rate has been derived 
in \cite{Rezzolla:2001di,Rezzolla:2001dh}, where it has been shown that
while the star remains in the instability region, the r-modes generate
a differential rotation which can greatly amplify a pre-existing magnetic
field. 
More specifically, if a poloidal magnetic field was originally present,
a strong toroidal field is generated inside the star.
The energy of the modes is therefore transferred to the magnetic field and
the instability is damped.
\\
We assume that the internal stellar magnetic field $\textbf{B}$ is initially
based on the solution obtained by Ferraro~\cite{Ferraro1954ApJ} which can be written \cite{Cuofano:2009yg}
\begin{eqnarray}
&&\mbox{\textbf{B}}_{0}^{in}(t=0)=\\ \nonumber
&& B_d \, \left[\left(-3\frac{r^2}{R^2}+5\right)
\texttt{cos} \, \theta \,\mbox{\textbf{e}}_r+\left(6\frac{r^2}{R^2}-5\right)\, \texttt{sin} \,
\theta \, \mbox{\textbf{e}}_\theta\right] \, .
\label{eq1M}
\end{eqnarray}
where $B_d$ is the strength of the equatorial magnetic field at the
stellar surface. \\
To estimate the magnetic field produced by r-modes we start by
writing the $l=m=2$ contribution to the perturbation velocity:
\begin{eqnarray}
\delta\textbf{v}(r,\theta,\phi,t)=\alpha\Omega
R\left(\frac{r}{R}\right)^2\textbf{Y}_{22}^Be^{i\sigma t}.
\end{eqnarray}
Following Ref.~\cite{Rezzolla:2001di} we get the total azimuthal displacement
from the onset of the oscillation at $t_0$ up to time $t$, which reads:
\begin{eqnarray}
&&\Delta \tilde{x}^{\phi}(r,t)\equiv \int_{t_0}^t \delta v^\phi(t')dt'  \nonumber \\
&& = \frac{2}{3}\left(\frac{r}{R}\right)
k_2(\theta)\int_{t_0}^t\alpha^2(t')\Omega(t')dt'+\mathcal{O}(\alpha^3) \,\,\,\,\,\,\,\,\,
\end{eqnarray}
where $k_2(\theta)\equiv (1/2)^7(5!/\pi)(\texttt{sin}^2\theta-2\texttt{cos}^2\theta)$.
The relation between the new and the original magnetic field inside the star 
in the Lagrangian approach reads~\cite{Rezzolla:2001di}:
\begin{eqnarray}
\frac{B^j}{\rho}(\tilde{\textbf{x}},t)=\frac{B^k}{\rho}(\textbf{x},t_0)
\frac{\partial\tilde{x_j}(t)}{\partial x^k(t_0)}.
\end{eqnarray}
This equation implies that the radial dependence of the initial and 
final magnetic field is the same.
Integrating on time the induction equation in the Eulerian approach one gets~\cite{Rezzolla:2001dh}:
\begin{eqnarray}
\delta B^{\theta} &\simeq& \delta B^r \simeq 0  \nonumber \\
\delta B^{\phi} &\simeq& B_0^{\theta} \int \dot{\phi}(t') dt' 
\simeq B_0^{\theta}\int\frac{\delta v^\phi(t')}{r}dt'
\label{eq2M}
\end{eqnarray}
where $B_{\phi}$ is the toroidal component.\\
The expression of the magnetic damping rate reads \cite{Cuofano:2009yg}:
\begin{eqnarray}
\frac{1}{\tau_{m_i}(t)} &=& \frac{(dE_M/dt)}{\tilde{E}} \nonumber \\ 
&\simeq& \frac{4 \int_0^{2\pi} d\phi \int_0^\pi  k_2^2(\theta) \texttt{sin}\theta d\theta 
\int_0^R r^4 B^2(r,\theta) dr}{9\pi \cdot (8.2\times 10^{-3}) M R^4 \Omega} \nonumber \\ 
& \times & \int^t_0 \alpha^2(t ')\Omega(t ') dt' \,\,\,\,\,\,\,\,\,\,\, \nonumber \\
&\simeq& \frac{4 A}{9\pi \cdot (8.2\times 10^{-3})}
\frac{B_d^2 R\int^t_0 \alpha^2(t ')\Omega(t ') dt '}{M\Omega}\,\,\,\,\,\,\,\,\,\,\,
\label{eq14}
\end{eqnarray}
where $\tilde{E}$ is the energy of the mode, $E_M$ is the magnetic
energy and $A\approx 0.99$. 
The time integral over the r-mode amplitude $\alpha$ takes contribution from the period
during which the star is inside the instability region.  

\subsection{Temperature evolution}
The equations derived above are strongly dependent on the value of the temperature, 
therefore we need to compute also the thermal evolution of the star. For the sake of simplicity
we assume that the temperature is uniform in the star and we make use of the estimates derived in Ref.~\cite{Andersson:2001ev}, for the case of
quark stars. The heat capacity reads:
\begin{eqnarray}
C_V=1.5\times 10^{38} M_{1.4}^{2/3} R_{10} T_8 \quad erg/K
\end{eqnarray}

The star cools down essentially by emitting neutrinos, through the so called 
URCA processes. Taking into account the direct URCA process, which are likely to occur
in the core of quark stars, the cooling rate due to emission of neutrinos reads: \cite{Andersson:2001ev}:
\begin{eqnarray}
\dot E_{neutrino}=3.77\times 10^{37} R_{10}^3 T_8^6 \quad erg/s
\end{eqnarray}
If the star is accreting mass then one has to consider the heating due
to the accreted material. In particular, for a quark star the heating
source comes from the conversion of nucleons into strange
matter. Estimating the energy release from this conversion to be
around 20 MeV per nucleon \cite{Alcock:1986hz,Andersson:2001ev}, the
heating rate due to the accretion is given by:
\begin{eqnarray}
\dot E_{accretion}=1.19\times 10^{37} \dot M_{-8} \quad erg/s
\end{eqnarray}
where $\dot M_{-8}$ is the accretion rate in units of $10^{-8}  M_\odot$ per year.
The last contribution to the heating of the star comes from the viscosity. Taking into account the contributions of both bulk and shear viscosity, the associated heating rate reads \cite{Owen:1998xg}:
\begin{eqnarray}
\dot E_{viscosity}=2 \alpha^2\Omega^2 M R^2 \tilde J \big(\frac{1}{\tau_{s}}+\frac{1}{\tau_b}\big) \quad erg/s
\end{eqnarray}
where the expression for $\tau_s$ and $\tau_b$ are given in eqs.~(\ref{svis}) and (\ref{bvis}), respectively, and $\tilde J=(2/3)\times 9.4\times 10^{-2}\tilde I$.

Finally, the equation for the thermal evolution of the star is given by:
\begin{eqnarray}
\frac{d}{dt}\big(\frac{1}{2} C_V T\big)=\dot E_{accretion}+ \dot E_{viscosity}
- \dot E_{neutrino}
\label{eq21}
\end{eqnarray}

In the absence of r-modes, the temperature is determined
by a balance between accretion heating and neutrino
cooling:
\begin{eqnarray}
T=8.25\times 10^7 \big(\dot M_{-8} R_{10}^{-3}\big)^{1/6} K
\label{eq22}
\end{eqnarray}

\section{Results and discussion}

By solving the Eqs.(\ref{eq4},\ref{eq5},\ref{eq14},\ref{eq21}) it is
possible to obtain the temporal evolution of quark stars, taking into
account the formation of toroidal magnetic fields due to r-mode
instability. In particular, in the present work we show the results
only for the case of quark stars inside of LMXBs. Eq.~(\ref{eq2M})
gives the strength of the new toroidal component created by the
winding up of a pre-existent poloidal field, due to the differential
rotation induced by r-modes.  Several issues remain open concerning
the way the new magnetic fields, produced by the damping of the r-modes, are
affected by possible instabilities. In the stably stratified
matter of a stellar interior there are two types of
instabilities: the Parker (or magnetic buoyancy) and the Tayler
instabilities (or pinch-type), both driven by the magnetic field
energy in the toroidal field.
The condition for the Tayler instability to set in is given by \cite{1999A&A...349..189S,2002A&A...381..923S}:
\begin{eqnarray}
\frac{\omega_A}{\Omega}>\big(\frac{N_{\mu}}{\Omega}\big)^{1/2}\big(\frac{\eta}{r^2\Omega}\big)^{1/4}\label{Tayler}
\end{eqnarray}
where $\omega_A=B/(4 \pi \rho)^{1/2} r$ is the Alfven frequency,
$N_{\mu}\simeq 5\times 10^4 s^{-1}$ is the compositional contribution
to the buoyancy frequency and $\eta\sim 10^{-9}f(m_s) T_8^2 cm^2
s^{-1}$ is the magnetic diffusivity which can be obtained from the
electrical conductivity $\sigma$ by using the relation
$\eta=1/(\mu_0\sigma)$ \cite{Haensel:1991pi} and $f(m_s)$ takes into
account the dependence of the electron fraction on the mass of the
strange quark and it ranges from $11$ to $4$ to $2.5$ when $m_s$
changes from $100$ to $200$ to $300$~MeV. From Eq.~(\ref{Tayler}) we
can conclude that in a pure quark star, the Tayler instability
sets in for $B> B_{cr} \sim 10^{12}$ G. After the development of the
Tayler instability, the toroidal component of the field produces, as a
result of its decay, a new poloidal component which can then be wound
up itself, closing the dynamo loop. When the differential rotation
stops the field can evolve into a stable configuration of a mixed
poloidal-toroidal twisted-torus shape inside the star with an
approximately dipolar field connected to it outside the star
\cite{2006A&A...449..451B,2004Natur.431..819B,2006A&A...450.1097B,2006A&A...450.1077B}.
Concerning the buoyancy instability, it occurs for $B_{cr}\gtrsim
10^{15}$ G \cite{Haensel:2007tf}, therefore it is not relevant in our
calculations.
\begin{figure}
    \begin{centering}
\epsfig{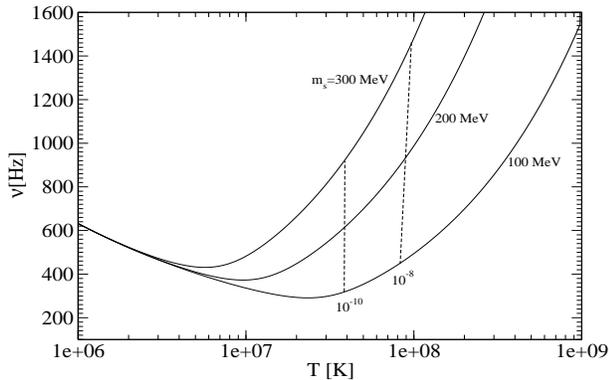}
    \caption{The quark star instability window for three values of the strange quark mass. The dashed lines 
indicate the evolution of a star in a LMXB during the accretion stage up to the moment in which 
the star enters the instability window.
\label{window} }
   \end{centering}
\end{figure}
\\ 
It is very important to point out that the formation and the further
stabilization of magnetic fields in the core of compact stars, leads
to different scenarios depending of the star composition.  For
instance, neutron stars contain a highly conductive crust acting as a
screen for the internal large magnetic field.  Conversely, whether
quark stars contain or not a crust is still controversial.  Various
types of crust have been suggested: a tenuous crust made of ions
suspended on the electric field associated with the most external
layer of a quark star \cite{Alcock:1986hz}; a crust made of a mixed
phase of electrons and of quark nuggets \cite{Jaikumar:2005ne}; a
crust made of a mixed phase of hadronic and quark matter
\cite{Drago:2001nq}. The electric properties of the crust area
strongly connected with the electron density, which is suppressed in
the core (what has been taken into account above, while discussing the
magnetic diffusivity) and which can be enhanced in the crust, at least
in the model in which a fraction of hadron is present in the crust
\cite{Drago:2001nq}. Although detailed calculations of the structure
of the crust of a quark star are not yet available, we can assume that
if a crust exists its electric properties interpolate between those of
a normal neutron star and those of a bare quark star, the two limiting
situations.
\\
As discussed above, when Tayler instability sets in, a poloidal
component with strength similar to the toroidal one ($B\sim 10^{12}$
G) is generated.  If the crust is not present (the case with a crust
will be shortly discussed later in the conclusions) such large
poloidal component is not screened and diffuses outside of the star,
preventing the star from further accreting mass. Therefore, following
this scenario, if a quark star inside of a LMXB produces, due to the
r-mode instability, an internal magnetic field larger than the Tayler
instability threshold, then it should stop accreting mass. Thus this
represents, at least in principle, an additional constraint on the
highest rotational frequency of the compact stars in LMXBs. \\
\begin{figure}
    \begin{centering}
\epsfig{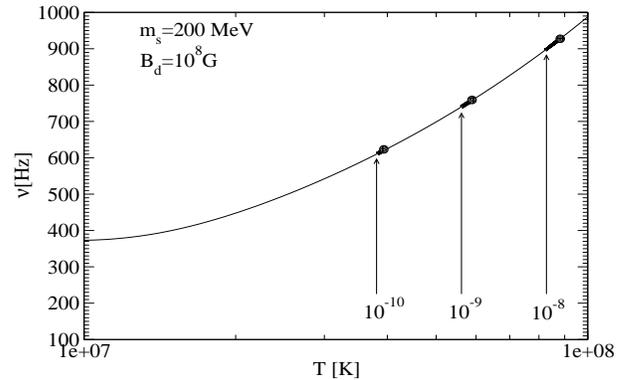}
    \caption{Instability window and trajectories of the evolution 
of the frequency and temperature of the star for different values of the mass accretion rate.
The short thick lines represent the paths followed by the star after the instability windows is touched.
The dots indicate the onset of the Tayler instability at which the evolution is stopped.
\label{tra200} }
   \end{centering}
\end{figure} 
There are many parameters that enter into the equations for the
temporal evolution of the frequency of the star.  We fix the mass and
the radius of the star to be $1.4 M_{\odot}$ and $10$ km
respectively. A crucial quantity for the bulk viscosity is the mass of
the strange quark for which we will consider the three values
$m_s=100-200-300$ MeV. The smallest value for $m_s$ is basically the
current mass of the strange quark as used within MIT bag-like models
and the last value is compatible with results of chiral quark models
as the NJL model. The astrophysical input parameters concern the mass
accretion rate, for which we use the values in the range $(10^{-10} -
10^{-8}) \,\mbox{M}_{\odot} \mbox{yr}^{-1}$, and the initial dipolar
magnetic field varied in the interval ${\bf B_{0}}=10^8 - 10^9$ G.

\begin{figure}
    \begin{centering}
\includegraphics*[height=4cm,width=8cm]{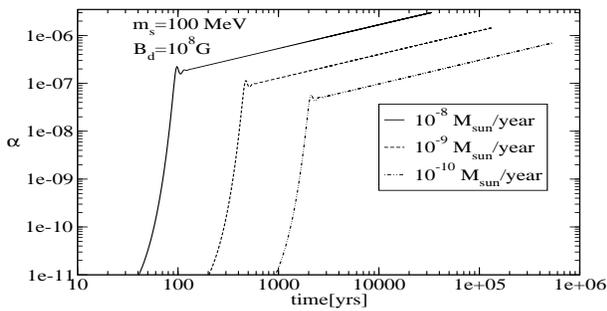}
    \caption{Temporal evolution of $\alpha$ for different values of the mass accretion rate. Here
the initial poloidal magnetic field is $B_d=10^8$~G and $m_s=100$~MeV.
\label{alpha100} }
   \end{centering}
\end{figure}

\begin{figure}
    \begin{centering}
\includegraphics*[height=4cm,width=8cm]{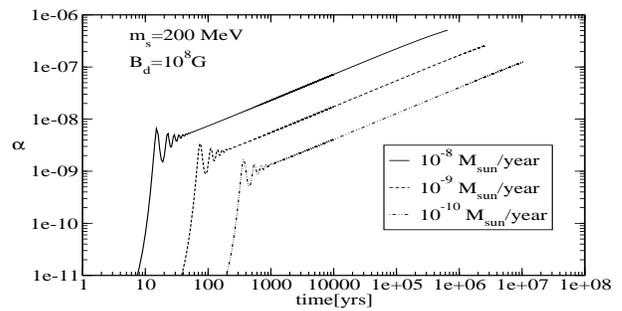}
    \caption{As in Fig.~\ref{alpha100} with $m_s=200$~MeV.
\label{alpha200}}
   \end{centering}
\end{figure}

Let us start our discussion with the r-modes instability window shown
in Fig.~\ref{window}.  With increasing values of $m_s$, the bulk
viscosity increases and therefore the instability region becomes
smaller.  Also shown by dashed lines are the trajectories of the
evolution of a star in a LMXB during the accretion stage and until the
star enters the instability window for two extreme values of the mass
accretion ($10^{-10}$ and $10^{-8}$ $M_{\odot} yr^{-1}$ ).
\\
The evolution is calculated by starting from a configuration below the
instability window: an initial value of $100-200$ Hz is taken as
initial frequency, $\alpha$ is set to zero and the initial temperature
is the equilibrium temperature as expressed by Eq.(\ref{eq22}).  Due
to accretion, the star is spun up until it reaches the instability
window and at that point the r-modes instability starts to
develop. The bulk viscosity dissipates part of the r-mode energy into
heat with a consequent reheating of the star and the internal magnetic
field starts to grow. At the same time, the frequency of the star
continues to grow due to the accretion torque, and the amplitude of
the r-modes increases. The simultaneous effect of reheating and
accretion leads the star to follow the border of the instability
window. This is clearly shown in Fig.~\ref{tra200}, where the thick
lines indicate the paths followed by the star (for different values of
the accretion rate) and the dots, at which the evolution stops,
signal the onset of the Tayler instability.  The temporal evolution of
the r-modes amplitude is shown in Fig.~\ref{alpha100} and
Fig.~\ref{alpha200}, where we take $m_s=100$ MeV and $m_s=200$ MeV,
respectively.  For both the cases the initial dipolar magnetic field
has a value of $10^{8}$ G.

\begin{figure}
    \begin{centering}
\includegraphics*[height=4cm,width=8cm]{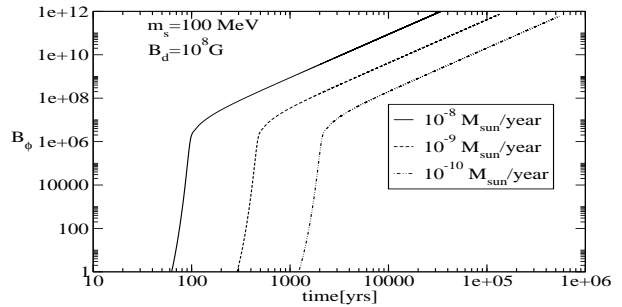}
    \caption{Temporal evolution of the internal toroidal magnetic field $B_{\varphi}$
 for different values of the mass accretion rate. Here $B_d=10^8$~G and $m_s=100$~MeV.
\label{btor100} }
   \end{centering}
\end{figure} 
\begin{figure}
    \begin{centering}
\includegraphics*[height=4cm,width=8cm]{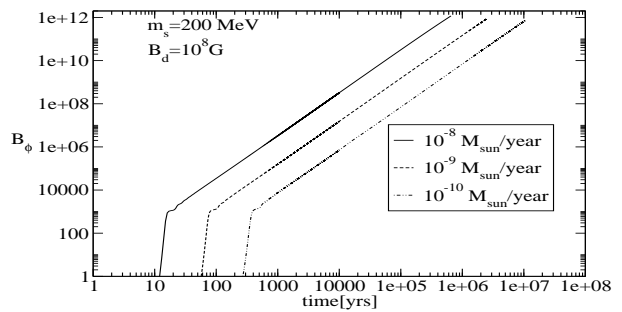}
    \caption{As in Fig.~\ref{btor100} with $m_s=200$~MeV.
\label{btor200} }
   \end{centering}
\end{figure} 
Notice that after an initial stage in which $\alpha$ oscillates (the star enters
and exits the instability region), $\alpha$ then increases steadily
but its value is still so small that r-modes cannot reach the
saturation regime. So our calculations are independent of the value of $\alpha$ in the saturation regime.
Parallel to the evolution of $\alpha$ we show in Figs.~\ref{btor100} and \ref{btor200}
the temporal evolution of the internal magnetic field. As expected, the value of $B_{\phi}$
follows the same evolution of $\alpha$. Again, we stopped the evolution as soon as $B_{\phi}$ 
reaches the threshold for the Tayler instability $B_{cr}\sim 10^{12}$ G.

\begin{figure}
    \begin{centering}
\includegraphics*[height=6cm,width=9cm]{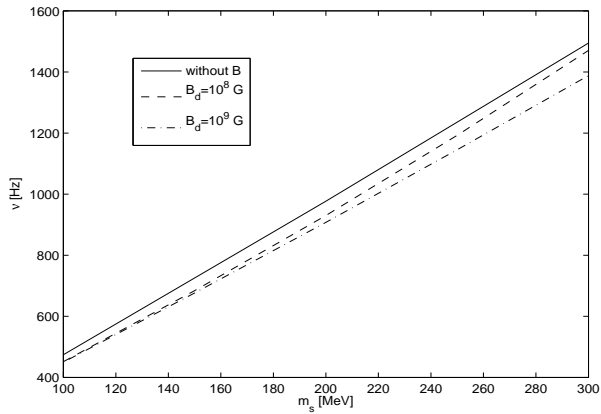}
    \caption{Maximum spin frequencies of quark stars as a function of the mass of the strange quark. 
The results obtained assuming a initial poloidal magnetic field $B_d=10^8$~G (dashed line) and $B_d=10^9$~G 
(dot-dashed line) are compared to the result obtained without the internal magnetic field (solid line).
\label{limiti} }
   \end{centering}
\end{figure}

\begin{figure}
    \begin{centering}
\includegraphics*[height=6cm,width=8cm]{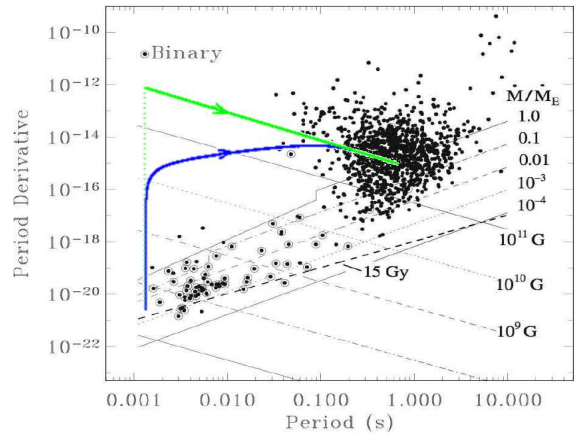}
    \caption{Trajectories in the plane $P\mbox{--}\dot{P}$ of quark stars after the development 
of the Tayler instability. The green line corresponds to a quark star without a crust; the blue 
line is obtained if in the crust the main diffusive process is the ambipolar diffusion.
\label{evolution} }
   \end{centering}
\end{figure}

As discussed in Ref.~\cite{Andersson:2001ev}, in which the effect of
internal magnetic damping was not considered, a quark star inside a
LMXB can be spun up up to a maximum frequency, corresponding to the
frequency at which the accretion torque balances the spin down torque
due to the emission of gravitational waves. However, taking into
account the internal magnetic damping, a quark star without a crust
should stop accreting after the onset of Tayler instability is reached. Therefore
an important question is how much the limiting frequency is reduced
with respect to the case in which the internal magnetic field is not taken
into account. To this purpose, we computed the maximum frequencies for
both the cases, estimating in this way the effect of the internal
magnetic field. These frequencies are shown in Fig.~\ref{limiti} as a
function of $m_s$ and for two values of the initial poloidal magnetic
field, $B_d=10^8$~G (dashed line) and $B_d=10^9$~G (dot-dashed line).
Notice that the effect of the internal magnetic field is rather small, reducing the maximum
rotational frequency only by a few tens Hertz. Moreover, the proven existence
of stars in LMXBs rotating at frequencies larger than $600$~Hz
rules out a value of $m_s\sim 100$~MeV.
\\
Finally, we investigate the possible evolutionary scenarios of quark
stars beyond the onset of the Tayler instability. Let us first
consider a star without a crust: as soon as the Tayler instability
sets in, the new magnetic configuration prevents the star from further
accreting mass. The new poloidal component, of the same order of
magnitude of the toroidal component ($\sim 10^{12}$ G), will act as a
strong braking torque, and the star will lose angular momentum. 
Such evolutionary path is plotted in
the plane $P\mbox{--}\dot{P}$ in Fig.(\ref{evolution}) and is depicted
by the green line. It is very interesting to notice that, starting
from the region of LMXBs, the star evolves into the region of radio
pulsars.

On the contrary, a highly conductive crust could screen the internal
magnetic field for a very long time. A possible example of such a
crust is that described in Ref.~\cite{Drago:2001nq} where the most
external layer is made of an admixture of hadrons and quarks. In that
case the electron fraction (and therefore also the electric
conductivity) is large in the crust and decreases towards to core of the star. 
The evolution of the star depends on the dominant dissipation
mechanism associated with the crust. The most relevant one is
probably the ambipolar diffusion (see
Refs.~\cite{Cuofano:2009xy,Cuofano:2009yg}), whose typical timescale
is:
\begin{eqnarray}
t_{ambip}\sim 3\times 10^9 \frac{T_8^2 L_5^2}{B_{12}^2} yr
\label{amb}
\end{eqnarray},
where $L_5=L/10^5$ cm is the size of the region embedding the magnetic field.
Due to this diffusion mechanism, the internal magnetic field can ``quite rapidly'' (in a few millions years) diffuse 
outside of the crust, and also in this case the
star should stop accreting. In the meanwhile, the external magnetic field should spin down the star.
In Fig.~\ref{evolution} we indicate with a blue line a sketch of the path followed by the star when 
the ambipolar diffusion is present; also in this case the star evolves into the region of the radio pulsars.

It is important to remark that the time spent by the quark star
after the Tayler instability and before it reaches the region
of the radio pulsars is of the order of a few million years, to be
compared with the time spent inside the radio pulsar region, which is
of the order of a few hundreds million years. Therefore we can expect
that only a few percent of the stars following the trajectory indicated
in Fig(\ref{evolution}) will be detected before reaching the radio pulsar region.

\section{Conclusions}
We have shown that strong toroidal fields can be generated in the core
of an accreting millisecond quark star which enters the r-mode
instability window. Tayler instability sets in when the generated
toroidal fields exceed the critical value $B_{tor}^{cr}\sim 10^{12}$~G
and a new poloidal component of similar strength is produced. Our
results show that the maximum spin frequency for quark stars does not
change significantly when taking into account the internal generated
magnetic fields. \\ The scenario after the development of the Tayler
instability depends on the presence and the properties of a possible
crust. If the crust is not present, the generated large poloidal
component diffuses quickly outside the core and prevents the further
accretion of mass on the star. On the other hand, if a highly
conductive crust is present, it could screen to some extent the
internal magnetic field. However, taking into account the ambipolar
diffusion, which is, in this case, the dominant dissipation mechanism,
the star could expel the internal magnetic field in a few millions
years, which would then stop the accretion. In both cases the quark
star evolves into the region of radio pulsars, as shown in
Fig.\ref{evolution}: this represents a new possible scenario
for the formation of radio pulsars.

\acknowledgments
L.B. is supported by CompStar a research program of the European
Science Foundation.  G.P. is supported by the German Research
Foundation (DFG) under Grant No. PA1780/2-1. J.~S.~B. is supported by
DFG through the Heidelberg Graduate School of Fundamental Physics.

\bibliography{references}
\bibliographystyle{apsrev4-1}

\end{document}